\documentclass[runningheads]{llncs}
\usepackage{booktabs}  
\usepackage{array}
\usepackage{bm}
\usepackage{multirow}
\usepackage{amssymb}
\usepackage{amsmath}
\usepackage{cite}
\usepackage{url}
\usepackage{bbding}
\usepackage{xspace}
\usepackage{graphicx}

\usepackage{mathrsfs}
\usepackage{algorithmic}
\usepackage[ruled,vlined]{algorithm2e}
\usepackage{epsfig}
\usepackage{textcomp}
\usepackage{diagbox}
\usepackage[dvipsnames]{xcolor}
\usepackage{wrapfig}
\usepackage[pagebackref=false,breaklinks=true,letterpaper=true,colorlinks,bookmarks=false]{hyperref}

\begin{document}
%
\title{Parse and Recall: Towards Accurate Lung Nodule Malignancy Prediction like Radiologists}
\titlerunning{Towards Accurate Lung Nodule Malignancy Prediction like Radiologists}

\author{Jianpeng Zhang\Envelope$^{1,4}$, 
Xianghua Ye\Envelope$^{2}$,
Jianfeng Zhang$^{1,4}$, 
Yuxing Tang$^{1}$,
Minfeng Xu$^{1,4}$,
Jianfei Guo$^{1,4}$,
Xin Chen$^{3}$,
Zaiyi Liu$^{3}$, \\
Jingren Zhou$^{1,4}$,
Le Lu$^{1}$,
Ling Zhang$^{1}$\\
}
\authorrunning{Zhang et al.}
\institute{$^1$ DAMO Academy, Alibaba Group
\\$^2$ The First Affiliated Hospital of College of Medicine, Zhejiang University, China
\\$^3$ Guangdong Provincial People’s Hospital, China
\\$^4$ Hupan Lab, 310023, Hangzhou, China
\\\email{jianpeng.zhang0@gmail.com; hye1982@zju.edu.cn}
}

\maketitle              
\begin{abstract}
Lung cancer is a leading cause of death worldwide and early screening is critical for improving survival outcomes. In clinical practice, the contextual structure of nodules and the accumulated experience of radiologists are the two core elements related to the accuracy of identification of benign and malignant nodules. Contextual information provides comprehensive information about nodules such as location, shape, and peripheral vessels, and experienced radiologists can search for clues from previous cases as a reference to enrich the basis of decision-making. 
In this paper, we propose a radiologist-inspired method to simulate the diagnostic process of radiologists, which is composed of context parsing and prototype recalling modules. 
The context parsing module first segments the context structure of nodules and then aggregates contextual information for a more comprehensive understanding of the nodule. 
The prototype recalling module utilizes prototype-based learning to condense previously learned cases as prototypes for comparative analysis, which is updated online in a momentum way during training. 
Building on the two modules, our method leverages both the intrinsic characteristics of the nodules and the external knowledge accumulated from other nodules to achieve a sound diagnosis. 
To meet the needs of both low-dose and noncontrast screening, we collect a large-scale dataset of 12,852 and 4,029 nodules from low-dose and noncontrast CTs respectively, each with pathology- or follow-up-confirmed labels. 
Experiments on several datasets demonstrate that our method achieves advanced screening performance on both low-dose and noncontrast scenarios.  

\end{abstract}

\section{Introduction}

Lung cancer screening has a significant impact on the rate of mortality associated with lung cancer. Studies have proven that regular lung cancer screening with low-dose computed tomography (LDCT) can lessen the rate of lung cancer mortality by up to 20\%~\cite{NLST2011,ardila2019end}. 
As most (e.g., 95\% \cite{mazzone2022evaluating}) of the detected nodules are benign, it is critical to accurately assess their malignancy on CT to achieve a timely diagnosis of malignant nodules and avoid unnecessary procedures such as biopsy for benign ones. Particularly, the evaluation of nodule (i.e., 8--30mm) malignancy is recommended in the guidelines \cite{mazzone2022evaluating}.

\begin{wrapfigure}{r}[0cm]{0pt}
\includegraphics[width=0.5\linewidth]{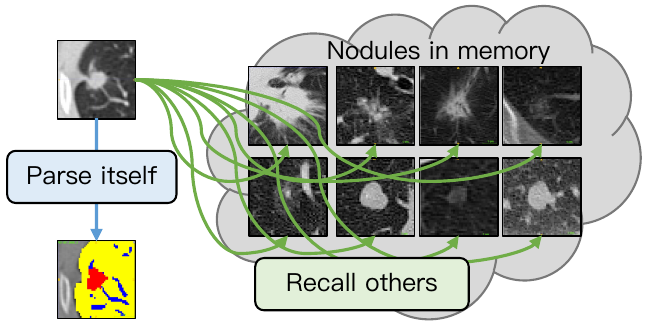}
\vspace{-1.5cm}
\caption{In PARE, a nodule is diagnosed from two levels: first \textbf{parsing} the contextual information contained in the nodule itself, and then \textbf{recalling} the previously learned nodules to look for related clues.  
}
\label{fig1}
\vspace{-0.5cm}
\end{wrapfigure}

One of the major challenges of lung nodule malignancy prediction is the quality of datasets~\cite{choi2022cirdataset}. It is characterized by a lack of standard-of-truth of labels for malignancy~\cite{LIDP_miccai2022,zhang2023trustworthy}, and due to this limitation, many studies use radiologists' subjective judgment on CT as labels, such as LIDC-IDRI~\cite{LIDC-IDRI}. 
Recent works have focused on collecting pathologically labeled data to develop reliable malignancy prediction models~\cite{wang2022deepln,LIDP_miccai2022,UnitedImaging_TMI2021}. For example, Shao et al.~\cite{LIDP_miccai2022} collated a pathological gold standard dataset of 990 CT scans. 
Another issue is most of the studies focus on LDCT for malignancy prediction~\cite{liao2019evaluate}. 
However, the majority of lung nodules are incidentally detected by routine imaging other than LDCT~\cite{bi2019artificial,osarogiagbon2022lung_JCO}, such as noncontrast chest CT (NCCT, the most frequently performed CT exam, nearly 40\% \cite{sodickson2009recurrent}). 

Technically, current studies on lung nodule malignancy prediction mainly focus on deep learning-based techniques~\cite{xie2017transferable,xie2018knowledge,liao2019evaluate,CA-Net_miccai2021,UnitedImaging_TMI2021}. 
Liao et al.~\cite{liao2019evaluate} trained a 3D region proposal network to detect suspicious nodules and then selected the top five to predict the probability of lung cancer for the whole CT scan, instead of each nodule.
To achieve the nodule-level prediction, Xie et al.~\cite{xie2018knowledge} introduced a knowledge-based collaborative model that hierarchically ensembles multi-view predictions at the decision level for each nodule. 
Liu et al.~\cite{CA-Net_miccai2021} extracted both nodules' and contextual features and fused them for malignancy prediction.
Shi et al.~\cite{UnitedImaging_TMI2021} effectively improved the malignancy prediction accuracy by using a transfer learning and semi-supervised strategy. 
Despite their advantages in representation learning, these methods do not take into account expert diagnostic knowledge and experience, which may lead to a bad consequence of poor generalization. 
We believe a robust algorithm should be closely related to the diagnosis experience of professionals, working like a radiologist rather than a black box. 

In this paper, we suggest mimicking radiologists' diagnostic procedures from intra-context \textbf{pa}rsing and inter-nodule \textbf{re}calling (see illustrations in Fig.~\ref{fig1}), abbreviated as \textbf{PARE}. 
At the intra-level, the contextual information of the nodules provides clues about their shape, size, and surroundings, and the integration of this information can facilitate a more reliable diagnosis of whether they are benign or malignant. 
Motivated by this, we first segment the context structure, \ie, nodule and its surroundings, 
and then aggregate the context information to the nodule representation via the attention-based dependency modeling, allowing for a more comprehensive understanding of the nodule itself. 
At the inter-level, we hypothesize that the diagnosis process does not have to rely solely on the current nodule itself, but can also find clues from past learned cases. This is similar to how radiologists rely on their accumulated experience in clinical practice. 
Thus, the model is expected to have the ability to store and recall knowledge, \ie, the knowledge learned can be recorded in time and then recalled as a reference for comparative analysis. 
To achieve this, we condense the learned nodule knowledge in the form of prototypes, 
and recall them to explore potential inter-level clues as an additional discriminant criterion for the new case. 
To fulfill both LDCT and NCCT screening needs, we curate a large-scale lung nodule dataset with pathology- or follow-up-confirmed benign/malignant labels.
For the LDCT, we annotate more than 12,852 nodules from 8,271 patients from the NLST dataset \cite{NLST2011}. 
For the NCCT, we annotate over 4,029 nodules from over 2,565 patients from our collaborating hospital. 
Experimental results on several datasets demonstrate that our method achieves outstanding performance on both LDCT and NCCT screening scenarios. 

Our contributions are summarized as follows:
(1) We propose context parsing to extract and aggregate rich contextual information for each nodule. 
(2) We condense the diagnostic knowledge from the learned nodules into the prototypes and use them as a reference to assist in diagnosing new nodules. 
(3) We curate the largest-scale lung nodule dataset with high-quality benign/malignant labels to fulfill both LDCT and NCCT screening needs. 
(4) Our method achieves advanced malignancy prediction performance in both screening scenarios (0.931 AUC), and exhibits strong generalization in external validation, setting a new state of the art on LUNGx (0.801 AUC).

\section{Method}
\begin{figure}[t]
	\begin{center}
		{\includegraphics[width=1.0\linewidth]{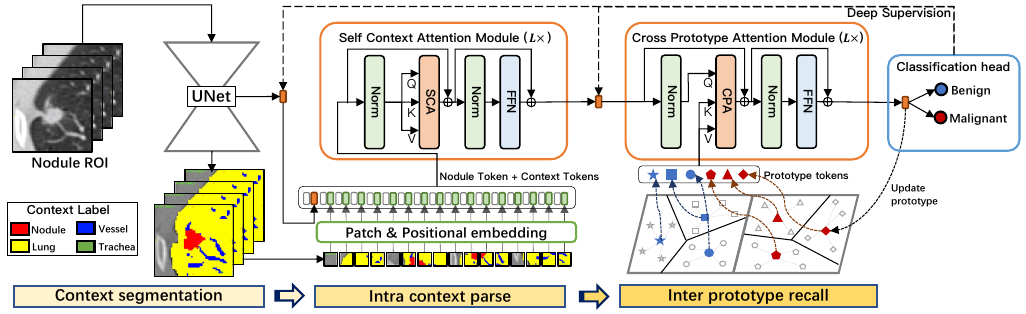}}
	\end{center}
 	\vspace{-0.7cm}
	\caption{Overview architecture of our proposed PARE model.
	}	
	\label{fig:framework}
	\vspace{-0.2cm}
\end{figure}

Fig.~\ref{fig:framework} illustrates the overall architecture of PARE, which consists of three stages: context segmentation, intra context parsing, and inter prototype recalling. We now delve into different stages in detail in the following subsections. 

\subsection{Context Segmentation}
The nodule context information has an important effect on the benign and malignant diagnosis. For example, a nodule associated with vessel feeding is more likely to be malignant than a solitary one~\cite{wu2010stratified}. Therefore, we use a U-like network (UNet) to parse the semantic mask $m$ for the input image patch $x$, thus allowing subsequent context modeling of both the nodule and its surrounding structures. Specifically, each voxel of $m$ belongs to $\{0:\rm background,1:\rm lung,2:\rm nodule,3:\rm vessel,4:\rm trachea\}$. This segmentation process allows PARE to gather comprehensive context information that is crucial for an accurate diagnosis. For the diagnosis purpose, we extract the global feature from the bottleneck of UNet as the nodule embedding $q$, which will be used in later diagnostic stages. 

\subsection{Intra Context Parse} 
In this stage, we attempt to enhance the discriminative representations of nodules by aggregating contextual information produced by the segmentation model. Specifically, the context mask is tokenized into a set of sequences via the overlapped patch embedding. 
The input image is also split into patches and then embedded into the context tokens to keep the original image information. 
Besides, positional encoding is added in a learnable manner to retain location information. 
Similar to the class token in ViT~\cite{dosovitskiyimage}, we prepend the nodule embedding token to the context sequences, denoted by $[q; t_1,...,t_g]\in \mathbb{R}^{(g+1)\times D}$. Here $g$ is the number of context tokens, and $D$ represents the embedding dimension. 
Then we perform the self-attention modeling on these tokens simultaneously, called Self Context Attention (SCA), to aggregate context information into the nodule embedding. 
The nodule embedding token at the output of the last SCA block serves as the updated nodule representation. We believe that explicitly modeling the dependency between nodule embedding and its contextual structure can lead to the evolution of more discriminative representations, thereby improving discrimination between benign and malignant nodules.

\subsection{Inter Prototype Recall}
\noindent\textbf{Definition of the prototype: } 
To retain previously acquired knowledge, a more efficient approach is needed instead of storing all learned nodules in memory, which leads to a waste of storage and computing resources. To simplify this process, we suggest condensing these pertinent nodules into a form of prototypes. 
As for a group of nodules, we cluster them into $N$ groups $\{C_1, ..., C_N\}$ by minimizing the objective function $\sum_{i=1}^N \sum_{p\in C_i} d(p, \bm{P}_i)$ where $d$ is the Euclidean distance function and $p$ represents the nodule embedding, and refer the center of each cluster, $\bm{P}_i = \frac{1}{|C_i|}\sum_{p\in C_i}p$, as its prototype. 
Considering the differences between benign and malignant nodules, we deliberately divide the prototypes into benign and malignant groups, denoted by $\bm{P}^B \in \mathbb{R}^{N/2\times D}$ and $\bm{P}^M \in \mathbb{R}^{N/2\times D}$. 

\noindent\textbf{Cross prototype attention:} 
In addition to parsing intra context, we also encourage the model to capture inter-level dependencies between nodules and external prototypes. This enables PARE to explore relevant identification basis beyond individual nodules. 
To accomplish this, we develop a Cross-Prototype Attention (CPA) module that utilizes nodule embedding as the query and the prototypes as the key and value. 
It allows the nodule embedding to selectively attend to the most relevant parts of prototype sequences. 
The state of query at the output of the last CPA module servers as the final nodule representation to predict its malignancy label, either ``benign" ($y=0$) or ``malignant" ($y=1$). 

\noindent\textbf{Updating prototype online:} The prototypes are updated in an online manner, thereby allowing them to adjust quickly to changes in the nodule representations. 
As for the nodule embedding $q$ of the data $(x, y)$, its nearest prototype is singled out and then updated by the following momentum rules,
\begin{equation}
\left\{\begin{matrix}
    \bm{P}_{{\arg\min}_j d(q, \bm{P}_j^B)}^B &= \lambda \cdot \bm{P}_{{\arg\min}_j d(q, \bm{P}_j^B)}^B + (1-\lambda)\cdot q & \; if \; y=0 \\
    \bm{P}_{{\arg\min}_j d(q, \bm{P}_j^M)}^M &= \lambda \cdot \bm{P}_{{\arg\min}_j d(q, \bm{P}_j^M)}^M + (1-\lambda)\cdot q & \; otherwise
\end{matrix}\right.
\label{Eq.update_prototype}
\end{equation}
where $\lambda$ is the momentum factor, set to 0.95 by default. The momentum updating can help accelerate the convergence and improve the generalization ability.

\subsection{Training Process of PARE}
The algorithm~\ref{algorithm1} outlines the training process of our PARE model which is based on two objectives: segmentation and classification. 
The Dice and cross-entropy loss are combined for segmentation, while cross-entropy loss is used for classification. 
Additionally, deep classification supervision is utilized to enhance the representation of nodule embedding in shallow layers like the output of the UNet and SCA modules.

\SetKwInOut{Require}{Require}
\SetKwInOut{Initialize}{Initialize}
\begin{algorithm}[t]
\small
\caption{Training process of PARE model. 
}
\begin{algorithmic}[1]
\FOR {\textit{iteration} = 1, 2, ...}
\STATE $\{x, m, y\} \leftarrow {\rm Sample}(D)$  \hfill $\triangleright$ Sample a data
\STATE $\{q, s\} \leftarrow {\rm UNet}(x)$ \hfill $\triangleright$ Infer UNet backbone
\STATE $z_0 \leftarrow [q; t_1,...,t_g]$ \hfill $\triangleright$ Patch embedding and positional encoding
\STATE $p_3 \leftarrow {\rm MLP}(q)$ \hfill $\triangleright$ Deep classification supervision
\FOR {\textit{l} = 1, ..., L}
\STATE $z'_l \leftarrow {\rm SCA}({\rm LN}(z_{l-1})) + z_{l-1}$ \hfill $\triangleright$ self context attention
\STATE $z_l \leftarrow {\rm MLP}({\rm LN}(z'_l)) + z'_l$
\ENDFOR
\STATE $p_2 \leftarrow {\rm MLP}(z_L^0)$ \hfill $\triangleright$ Deep classification supervision
\FOR {\textit{l} = 1, ..., L}
\STATE $z'_l \leftarrow {\rm CPA}({\rm LN}(z_{l-1})) + z_{l-1}$ \hfill $\triangleright$ Cross prototype attention
\STATE $z_l \leftarrow {\rm MLP}({\rm LN}(z'_l)) + z'_l$
\ENDFOR
\STATE $p_1 \leftarrow {\rm MLP}(z_L^0)$ \hfill $\triangleright$ Classification head
\STATE Update prototype according to Eq.~\ref{Eq.update_prototype}
\STATE $J \leftarrow {\rm seg{\_}loss}(m, s) + \sum_{i=1}^3 {\rm cls{\_}loss}(y, p_i)$ \hfill $\triangleright$ Update loss
\ENDFOR 
\end{algorithmic}
\label{algorithm1}
\end{algorithm}

\section{Experiment}
\subsection{Datasets and Implementation Details}
\noindent\textbf{Data collection and curation:}
\textbf{NLST} is the first large-scale LDCT dataset for low-dose CT lung cancer screening purpose~\cite{NLST2011}. There are 8,271 patients enrolled in this study. 
An experienced radiologist chose the last CT scan of each patient, and localized and labeled the nodules in the scan as benign or malignant based on the rough candidate nodule location and whether the patient develops lung cancer provided by NLST metadata. 
The nodules with a diameter smaller than 4mm were excluded. 
\textbf{The in-house cohort} was retrospectively collected from 2,565 patients at our collaborating hospital between 2019 and 2022. Unlike NLST, this dataset is noncontrast chest CT, which is used for routine clinical care. 
\textbf{Segmentation annotation}: 
We provide the segmentation mask for our in-house data, but not for the NLST data considering its high cost of pixel-level labeling. 
The nodule mask of each in-house data was manually annotated with the assistance of CT labeler~\cite{wang2023cascaded} by our radiologists, while other contextual masks such as lung, vessel, and trachea were generated using the TotalSegmentator~\cite{totalsegmentator}.

\noindent\textbf{Train-Val-Test}: 
The training set contains 9,910 (9,413 benign and 497 malignant) nodules from 6,366 patients at NLST, and 2,592 (843 benign and 1,749 malignant) nodules from 2,113 patients at the in-house cohort. 
The validation set contains 1,499 (1,426 benign and 73 malignant) nodules from 964 patients at NLST. 
The NLST test set has 1,443 (1,370 benign and 73 malignant) nodules from 941 patients.
The in-house test set has 1,437 (1,298 benign and 139 malignant) nodules from 452 patients. 
We additionally evaluate our method on the LUNGx~\cite{armato2016lungx} challenge dataset, which is usually used for external validation in previous work~\cite{xie2018knowledge,liao2022learning,choi2022cirdataset}. LUNGx contains 83 (42 benign and 41 malignant) nodules, part of which (13 scans) were contrast-enhanced. 
\textbf{Segmentation}: We also evaluate the segmentation performance of our method on the public nodule segmentation dataset LIDC-IDRI~\cite{LIDC-IDRI}, which has 2,630 nodules with nodule segmentation mask. 
\textbf{Evaluation metrics}: The area under the receiver operating characteristic curve (AUC) is used to evaluate the malignancy prediction performance. 

\noindent\textbf{Implementation}: 
All experiments in this work were implemented based on the nnUnet framework~\cite{isensee2021nnu}, with the input size of $32\times 48\times48$, batch size of 64, and total training iterations of $10K$. 
In the context patch embedding, each patch token is generated from a window of $8\times 8\times 8$. 
The hyper-parameters of PARE are empirically set based on the ablation experiments on the validation set. For example, the Transformer layer is set to 4 in both SCA and CPA modules, and the number of prototypes is fixed to 40 by default. More details can be found in the ablation. 
Due to the lack of manual annotation of nodule masks for the NLST dataset, we can only optimize the segmentation task using our in-house dataset, which has manual nodule masks.

\subsection{Experiment results}
\noindent\textbf{Ablation study}: In Table~\ref{tab:ablation-1}, we investigate the impact of different configurations on the performance of PARE on the validation set, including Transformer layers, number of prototypes, embedding dimension, and deep supervision. We observe that a higher AUC score can be obtained by increasing the number of Transformer layers, increasing the number of prototypes, doubling the channel dimension of token embeddings, or using deep classification supervision. Based on the highest AUC score of 0.931, we empirically set L=4, N=40, D=256, and DS=True in the following experiments. 
In Table~\ref{tab:ablation-2}, we investigate the ablation study of different methods/modules on the validation set and observe the following results: 
(1) The pure segmentation method performs better than the pure classification method, primarily because it enables greater supervision at the pixel level, 
(2) the joint segmentation and classification is superior to any single method, indicating the complementary effect of both tasks, 
(3) both context parsing and prototype comparing contribute to improved performance on the strong baseline, demonstrating the effectiveness of both modules, 
and (4) segmenting more contextual structures such as vessels, lungs, and trachea provide a slight improvement, compared to solely segmenting nodules.

\begin{minipage}{\textwidth}
\begin{minipage}[t]{0.48\textwidth}
\makeatletter\def\@captype{table}
\small
\centering
\caption{Ablation comparison of hyper-parameters (Transformer layers (L), number of prototypes (N), embedding dimension (D), and deep supervision (DS))}
\begin{tabular}{p{0.8cm}<{\centering}|p{0.8cm}<{\centering}|p{0.8cm}<{\centering}|p{0.8cm}<{\centering}|p{1.2cm}<{\centering}}
    \toprule
    L & N & D & DS & AUC \\
    \midrule
    1 & 20 & 128 & $\checkmark$ & 0.912 \\
    2 & 20 & 128 & $\checkmark$ & 0.918 \\
    4 & 20 & 128 & $\checkmark$ & 0.924 \\
    4 & 10 & 128 & $\checkmark$ & 0.920 \\
    4 & 40 & 128 & $\checkmark$ & 0.924 \\
    4 & 40 & 256 & $\checkmark$ & \textbf{0.931} \\
    4 & 40 & 256 & $\times$ & 0.926 \\
    \bottomrule
\end{tabular}
\label{tab:ablation-1}
\end{minipage}
\begin{minipage}[t]{0.48\textwidth}
\makeatletter\def\@captype{table}
\small
\centering
\caption{Effectiveness of different modules. MT: multi-task learning. Context: intra context parsing. Prototype: inter prototype recalling. *: only nodule mask was used in the segmentation task.}
\begin{tabular}{c|c}
    \toprule
    Method      & AUC \\
    \midrule
    Pure classification & 0.907 \\
    Pure segmentation & 0.915 \\
    MT & 0.916 \\
    MT+Context* & 0.921 \\
    MT+Context & 0.924 \\
    MT+Context+Prototype &  \textbf{0.931}\\
    \bottomrule
\end{tabular}
\label{tab:ablation-2}
\end{minipage}
\end{minipage}

\noindent\textbf{Comparison to other methods on both screening scenarios}: 
Table~\ref{tab:sota} presents a comparison of PARE with other advanced methods, including pure classification-based, pure segmentation-based, and multi-task-based methods. 
Stratification assessments were made in both test sets based on the nodule size distribution. 
The results indicate that the segmentation-based method outperforms pure classification methods, mainly due to its superior ability to segment contextual structures. 
Additionally, the multi-task-based CA-Net outperforms any single-task method. Our PARE method surpasses all other methods on both NLST and In-house test sets. 
Moreover, by utilizing the ensemble of multiple deep supervision heads, the overall AUC is further improved to 0.931 on both datasets.

\begin{table}[]
\caption{Comparison of different methods on both NLST and in-house test sets. $^\dagger$: pure classification; $^\ddagger$: pure segmentation; $^\diamond$: multi-task learning; *: ensemble of deep supervision heads. Note that we add the segmentation task in CA-Net.}
\begin{tabular}{c|cccc|cccc}
\hline
\multirow{2}{*}{Method}    & \multicolumn{4}{c|}{NLST test set}                                                                                           & \multicolumn{4}{c}{In-house test set}                                                                                         \\ \cline{2-9} 
                           & \multicolumn{1}{c|}{\textless{}10mm} & \multicolumn{1}{c|}{10$\sim$20mm} & \multicolumn{1}{c|}{\textgreater{}20mm} & All   & \multicolumn{1}{c|}{\textless{}10mm} & \multicolumn{1}{c|}{10$\sim$20mm} & \multicolumn{1}{c|}{\textgreater{}20mm} & All   \\ \hline
CNN$^\dagger$ & \multicolumn{1}{c|}{0.742}           & \multicolumn{1}{c|}{0.706}        & \multicolumn{1}{c|}{0.780}               & 0.894 & \multicolumn{1}{c|}{0.851}           & \multicolumn{1}{c|}{0.797}        & \multicolumn{1}{c|}{0.744}              & 0.901 \\ \hline
ASPP~\cite{chen2017deeplab}$^\dagger$ & \multicolumn{1}{c|}{0.798}           & \multicolumn{1}{c|}{0.716}        & \multicolumn{1}{c|}{0.801}              & 0.902 & \multicolumn{1}{c|}{0.854}           & \multicolumn{1}{c|}{0.788}        & \multicolumn{1}{c|}{0.743}              & 0.901 \\ \hline
MiT~\cite{xie2022unimiss}$^\dagger$ & \multicolumn{1}{c|}{0.821}                & \multicolumn{1}{c|}{0.755}             & \multicolumn{1}{c|}{0.810}                   &  0.908     & \multicolumn{1}{c|}{0.858}                & \multicolumn{1}{c|}{0.784}             & \multicolumn{1}{c|}{0.751}                   &  0.904     \\ \hline
nnUnet~\cite{isensee2021nnu}$^\ddagger$ & \multicolumn{1}{c|}{0.815}           & \multicolumn{1}{c|}{0.736}        & \multicolumn{1}{c|}{0.815}              & 0.910  & \multicolumn{1}{c|}{0.863}           & \multicolumn{1}{c|}{0.804}        & \multicolumn{1}{c|}{0.750}               & 0.911 \\ \hline
CA-Net~\cite{CA-Net_miccai2021}$^\diamond$ & \multicolumn{1}{c|}{0.833}           & \multicolumn{1}{c|}{0.759}        & \multicolumn{1}{c|}{0.807}              & 0.916 & \multicolumn{1}{c|}{0.878}           & \multicolumn{1}{c|}{0.786}        & \multicolumn{1}{c|}{0.779}              & 0.918 \\ \hline
PARE$^\diamond$        & \multicolumn{1}{c|}{0.882}           & \multicolumn{1}{c|}{0.770}         & \multicolumn{1}{c|}{0.826}              & 0.928 & \multicolumn{1}{c|}{0.892}           & \multicolumn{1}{c|}{0.817}        & \multicolumn{1}{c|}{\textbf{0.783}}              & 0.927 \\ \hline
PARE$^\diamond$*          & \multicolumn{1}{c|}{\textbf{0.890}}            & \multicolumn{1}{c|}{\textbf{0.781}}        & \multicolumn{1}{c|}{\textbf{0.827}}              & \textbf{0.931} & \multicolumn{1}{c|}{\textbf{0.899}}           & \multicolumn{1}{c|}{\textbf{0.821}}        & \multicolumn{1}{c|}{0.780}               & \textbf{0.931} \\ \hline
\end{tabular}
\label{tab:sota}
\end{table}

\noindent\textbf{External evaluation on LUNGx}: 
We used LUNGx~\cite{armato2016lungx} as an external test to evaluate the generalization of PARE. It is worth noting that these compared methods have never been trained on LUNGx. Table~\ref{tab:external_LUNGx} shows that our PARE model achieves the highest AUC of 0.801, which is 2\% higher than the best method DAR~\cite{liao2022learning}. We also conducted a reader study to compare PARE with two experienced radiologists, who have 8 and 13 years of lung nodule diagnosis experience respectively. Results in Figure~\ref{fig:reader_study} reveal that our method achieves performance comparable to that of radiologists. 

\makeatletter\def\@captype{table}\makeatother
\begin{minipage}{.45\textwidth}
\small
\centering
\caption{Comparison to other competitive methods on LUNGx~\cite{armato2016lungx}.}
\begin{tabular}{l|l}
\toprule
Method           & AUC   \\ 
\midrule
NLNL~\cite{kim2019nlnl}  & 0.683 \\ 
D2CNN~\cite{yao2020deep} & 0.746 \\ 
KBC~\cite{xie2018knowledge}    & 0.768 \\ 
DAR~\cite{liao2022learning}    & 0.781 \\ 
PARE (Ours)             & \textbf{0.801} \\ 
\bottomrule
\end{tabular}
\label{tab:external_LUNGx}
\end{minipage}
\makeatletter\def\@captype{figure}\makeatother
\begin{minipage}{.5\textwidth}
\centering
\includegraphics[width=0.8\linewidth]{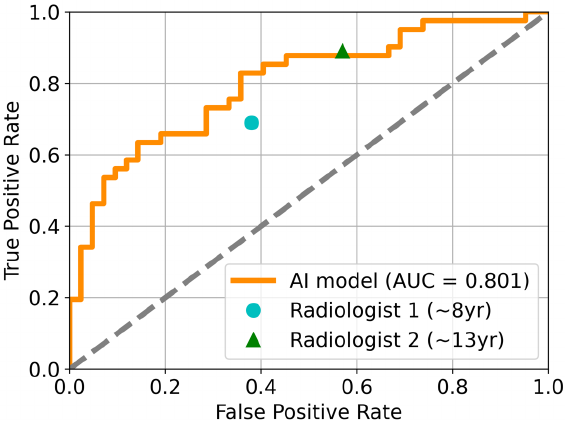}
\vspace{-0.5cm}
\caption{Reader study compared with AI.}
\label{fig:reader_study}
\end{minipage}

\vspace{0.5cm}
\noindent\textbf{Generalization on LDCT and NCCT}: 
Our model is trained on a mix of LDCT and NCCT datasets, which can perform robustly across low-dose and regular-dose applications. We compare the generalization performance of the models obtained under three training data configurations (LDCT, NCCT, and a combination of them). 
We find that the models trained on either LDCT or NCCT dataset alone cannot generalize well to other modalities, with at least a 6\% AUC drop. However, our mixed training approach performs best on both LDCT and NCCT with almost no performance degradation.

\section{Conclusion}
In this work, we propose the PARE model to mimic the radiologists' diagnostic procedures for accurate lung nodule malignancy prediction. 
Concretely, we achieve this purpose by parsing the contextual information from the nodule itself and recalling the previous diagnostic knowledge to explore related benign or malignancy clues. 
Besides, we curate a large-scale pathological-confirmed dataset with up to 13,000 nodules to fulfill the needs of both LDCT and NCCT screening scenarios. 
With the support of a high-quality dataset, our PARE achieves outstanding malignancy prediction performance in both scenarios and demonstrates a strong generalization ability on the external validation.


{
	\bibliographystyle{splncs04}
	\bibliography{egbib}

\begin{thebibliography}{10}
\providecommand{\url}[1]{\texttt{#1}}
\providecommand{\urlprefix}{URL }
\providecommand{\doi}[1]{https://doi.org/#1}

\bibitem{ardila2019end}
Ardila, D., Kiraly, A.P., Bharadwaj, S., Choi, B., Reicher, J.J., Peng, L.,
  Tse, D., Etemadi, M., Ye, W., Corrado, G., et~al.: End-to-end lung cancer
  screening with three-dimensional deep learning on low-dose chest computed
  tomography. Nat. Med.  \textbf{25}(6),  954--961 (2019)

\bibitem{armato2016lungx}
Armato~III, S.G., Drukker, K., Li, F., et~al.: {LUNGx} challenge for
  computerized lung nodule classification. J. Med. Imaging  \textbf{3}(4),
  044506--044506 (2016)

\bibitem{LIDC-IDRI}
Armato~III, S.G., McLennan, G., Bidaut, L., et~al.: The lung image database
  consortium ({LIDC}) and image database resource initiative ({IDRI}): a
  completed reference database of lung nodules on {CT} scans. Med. Phys.
  \textbf{38}(2),  915--931 (2011)

\bibitem{bi2019artificial}
Bi, W.L., Hosny, A., Schabath, M.B., Giger, M.L., Birkbak, N.J., Mehrtash, A.,
  Allison, T., Arnaout, O., Abbosh, C., Dunn, I.F., et~al.: Artificial
  intelligence in cancer imaging: clinical challenges and applications. CA: A
  Cancer Journal for Clinicians  \textbf{69}(2),  127--157 (2019)

\bibitem{chen2017deeplab}
Chen, L.C., Papandreou, G., Kokkinos, I., Murphy, K., Yuille, A.L.: Deeplab:
  Semantic image segmentation with deep convolutional nets, atrous convolution,
  and fully connected crfs. IEEE Trans. Pattern Anal. Mach. Intell.
  \textbf{40}(4),  834--848 (2017)

\bibitem{choi2022cirdataset}
Choi, W., Dahiya, N., Nadeem, S.: Cirdataset: A large-scale dataset for
  clinically-interpretable lung nodule radiomics and malignancy prediction. In:
  MICCAI. Springer (2022)

\bibitem{dosovitskiyimage}
Dosovitskiy, A., Beyer, L., Kolesnikov, A., et~al.: An image is worth 16x16
  words: Transformers for image recognition at scale. In: ICLR (2021)

\bibitem{isensee2021nnu}
Isensee, F., Jaeger, P.F., Kohl, S.A., Petersen, J., Maier-Hein, K.H.: nnu-net:
  a self-configuring method for deep learning-based biomedical image
  segmentation. Nat. Methods  \textbf{18}(2),  203--211 (2021)

\bibitem{kim2019nlnl}
Kim, Y., Yim, J., Yun, J., Kim, J.: {NLNL}: Negative learning for noisy labels.
  In: ICCV. pp. 101--110 (2019)

\bibitem{liao2019evaluate}
Liao, F., Liang, M., Li, Z., Hu, X., Song, S.: Evaluate the malignancy of
  pulmonary nodules using the {3D} deep leaky noisy-or network. IEEE Trans.
  Neural Netw. Learn. Syst.  \textbf{30}(11),  3484--3495 (2019)

\bibitem{liao2022learning}
Liao, Z., Xie, Y., Hu, S., Xia, Y.: Learning from ambiguous labels for lung
  nodule malignancy prediction. IEEE Trans. Med. Imaging  \textbf{41}(7),
  1874--1884 (2022)

\bibitem{CA-Net_miccai2021}
Liu, M., Zhang, F., Sun, X., et~al.: Ca-net: Leveraging contextual features for
  lung cancer prediction. In: MICCAI. pp. 23--32. Springer (2021)

\bibitem{mazzone2022evaluating}
Mazzone, P.J., Lam, L.: Evaluating the patient with a pulmonary nodule: a
  review. JAMA  \textbf{327}(3),  264--273 (2022)

\bibitem{NLST2011}
{National Lung Screening Trial Research Team}: Reduced lung-cancer mortality
  with low-dose computed tomographic screening. N. Engl. J. Med.
  \textbf{365}(5),  395--409 (2011)

\bibitem{osarogiagbon2022lung_JCO}
Osarogiagbon, R.U., Liao, W., Faris, N.R., Meadows-Taylor, M., Fehnel, C.,
  Lane, J., Williams, S.C., Patel, A.A., Akinbobola, O.A., Pacheco, A., et~al.:
  Lung cancer diagnosed through screening, lung nodule, and neither program: a
  prospective observational study of the detecting early lung cancer ({DELUGE})
  in the mississippi delta cohort. J. Clin. Oncol.  \textbf{40}(19), ~2094
  (2022)

\bibitem{LIDP_miccai2022}
Shao, Y., Wang, M., Mai, J., Fu, X., Li, M., Zheng, J., Diao, Z., Yin, A.,
  Chen, Y., Xiao, J., et~al.: {LIDP}: A lung image dataset with pathological
  information for lung cancer screening. In: MICCAI. pp. 770--779. Springer
  (2022)

\bibitem{UnitedImaging_TMI2021}
Shi, F., Chen, B., Cao, Q., Wei, Y., Zhou, Q., Zhang, R., Zhou, Y., Yang, W.,
  Wang, X., Fan, R., et~al.: Semi-supervised deep transfer learning for
  benign-malignant diagnosis of pulmonary nodules in chest {CT} images. IEEE
  Trans. Med. Imaging  \textbf{41}(4),  771--781 (2021)

\bibitem{sodickson2009recurrent}
Sodickson, A., Baeyens, P.F., Andriole, K.P., Prevedello, L.M., Nawfel, R.D.,
  Hanson, R., Khorasani, R.: {Recurrent CT, cumulative radiation exposure, and
  associated radiation-induced cancer risks from CT of adults}. Radiology
  \textbf{251}(1),  175--184 (2009)

\bibitem{wang2022deepln}
Wang, C., Shao, J., Xu, X., Yi, L., Wang, G., Bai, C., Guo, J., He, Y., Zhang,
  L., Yi, Z., et~al.: Deepln: A multi-task ai tool to predict the imaging
  characteristics, malignancy and pathological subtypes in ct-detected
  pulmonary nodules. Frontiers in Oncology  \textbf{12} (2022)

\bibitem{wang2023cascaded}
Wang, F., Cheng, C.T., Peng, C.W., Yan, K., Wu, M., Lu, L., Liao, C.H., Zhang,
  L.: A cascaded approach for ultraly high performance lesion detection and
  false positive removal in liver ct scans. arXiv preprint arXiv:2306.16036
  (2023)

\bibitem{totalsegmentator}
Wasserthal, J., Meyer, M., Breit, H.C., Cyriac, J., Yang, S., Segeroth, M.:
  Totalsegmentator: robust segmentation of 104 anatomical structures in {CT}
  images. arXiv preprint arXiv:2208.05868  (2022)

\bibitem{wu2010stratified}
Wu, D., Lu, L., Bi, J., Shinagawa, Y., Boyer, K., Krishnan, A., Salganicoff,
  M.: Stratified learning of local anatomical context for lung nodules in {CT}
  images. In: CVPR. pp. 2791--2798. IEEE (2010)

\bibitem{xie2017transferable}
Xie, Y., Xia, Y., Zhang, J., Feng, D.D., Fulham, M., Cai, W.: Transferable
  multi-model ensemble for benign-malignant lung nodule classification on chest
  ct. In: MICCAI. pp. 656--664. Springer (2017)

\bibitem{xie2018knowledge}
Xie, Y., Xia, Y., Zhang, J., Song, Y., Feng, D., Fulham, M., Cai, W.:
  Knowledge-based collaborative deep learning for benign-malignant lung nodule
  classification on chest {CT}. IEEE Trans. Med. Imaging  \textbf{38}(4),
  991--1004 (2018)

\bibitem{xie2022unimiss}
Xie, Y., Zhang, J., Xia, Y., Wu, Q.: Unimiss: Universal medical self-supervised
  learning via breaking dimensionality barrier. In: ECCV. pp. 558--575.
  Springer (2022)

\bibitem{yao2020deep}
Yao, Y., Deng, J., Chen, X., Gong, C., Wu, J., Yang, J.: Deep discriminative
  cnn with temporal ensembling for ambiguously-labeled image classification.
  In: AAAI. vol.~34, pp. 12669--12676 (2020)

\bibitem{zhang2023trustworthy}
Zhang, H., Chen, L., Gu, X., et~al.: Trustworthy learning with (un)sure
  annotation for lung nodule diagnosis with {CT}. Med. Image Anal.
  \textbf{83},  102627 (2023)

\end{thebibliography}
}

\newpage

\end{document}